\documentstyle[prl,twocolumn,aps,rotate,epsf,floats,amssymb,amsmath]{revtex}

\newcommand{\bn}{{\boldsymbol n}}
\newcommand{\bq}{{\boldsymbol q}}
\newcommand{\bk}{{\boldsymbol k}}

\newcommand{\bG}{{\boldsymbol G}}

\newcommand{\bP}{{\boldsymbol P}}
\newcommand{\bR}{{\boldsymbol R}}

\newcommand{\br}{{\boldsymbol r}}

\newcommand{\bDelta}{{\boldsymbol{\Delta}}}

\newcommand{\ox}{{\overline{x}}}
\newcommand{\oy}{{\overline{y}}}

\begin{document}
\setlength{\unitlength}{1cm}
\renewcommand{\arraystretch}{1.4}

\title{Orbital-singlet pairing and order parameter symmetry in Sr$_2$RuO$_4$}  
\author{Ralph Werner}

\address{Institut f\"ur Theorie der Kondensierten Materie,
  Universit\"at Karlsruhe, 76128 Karlsruhe, Germany}

\date{\today}

\maketitle

\centerline{Preprint. Typeset using REV\TeX}

\begin{abstract} 
Based on the degeneracy of the $d_{zx}$ and $d_{yz}$ orbitals in
Sr$_2$RuO$_4$ it is argued that the Cooper pairs condense in orbital
singlets. Together with the spin-triplet wave functions the real-space
wave function then is symmetric. Considering interaction effects the
order parameter is found to have $A_{1g}$ symmetry consistent with a
number of experimental observations. The sensitivity of the material
on non-magnetic impurities follows in a straightforward manner from
the orbital-singlet configuration.   
\end{abstract}


With the discovery of the high temperature superconductors a whole
class of transition metal oxides became a focal point in condensed
matter research. These materials exhibit many unconventional
properties whose interpretation has so far generally proved
controversial. An example that attracted a lot of attention is
Sr$_2$RuO$_4$. Its normal state properties are Fermi liquid like in
the temperature range $T_{\rm c} < T < 30$ K\cite{MYH+97,IMK+00} but
below $T_{\rm c} \le 1.5$ K the material is an unconventional
superconductor\cite{MRS01} since a number of experimental
probes\cite{IMK+98,LFK+98,DHM+00} show that the paired electrons carry
a magnetic moment. In spite of the large interest that the
superconductivity in Sr$_2$RuO$_4$ has attracted an unambiguous
understanding of the electronic correlations has not yet evolved. 

Rice and Sigrist\cite{RS95} proposed that the superconducting order
parameter has $p$-wave symmetry promoted by ferromagnetic correlations
by analogy with $^3$He. This idea is supported by experiments that show
that the static magnetic properties of Sr$_2$RuO$_4$ are the same in
the normal and the superconducting
phase\cite{IMK+98,DHM+00}. However, there is no conclusive
experimental proof\cite{LGL+00,TSN+01,ITY+01} for the $p$-wave
symmetry of the superconducting order parameter and no indications of
ferromagnetic correlations have been found either in neutron
scattering investigations\cite{SBB+99} or other
approaches\cite{MPS00,DLS+00,SDL+01}.

Furthermore, the specific heat\cite{NMM00}, nuclear quadrupole
resonance (NQR)\cite{IMK+00}, and  thermal conductivity\cite{STK+02}
are consistent with two-dimensional gapless fluctuations in the
superconducting phase of Sr$_2$RuO$_4$, which are incompatible with the
analogy to superfluid $^3$He. One possible scenario is the existence
of line nodes similar to those in the superconducting
cuprates\cite{TK00}. Since vertical line nodes have been ruled out by
thermal conductivity measurements\cite{TSN+01,ITY+01} horizontal line
nodes in the subsystem of the $d_{zx}$ and $d_{yz}$ electrons have
been proposed\cite{ZR01}. One weakness of the latter picture is that
it requires the fine tuning of various interaction
strengths\cite{KS02}, while no double gap structures have been observed
in Andreev reflection spectroscopy data\cite{LGL+00}. 

In this letter it is shown how the degeneracy of the Ru$^{4+}$
$d_{zx}$ and $d_{yz}$ orbitals allows for a straightforward
description of the unconventional superconductivity in Sr$_2$RuO$_4$
that is consistent with the experimental observations. The possibility
of mixed orbital pairing leading to $S=1$ spin-triplet Cooper pairs
through Hund's rule coupling has been raised implicitly by
Baskaran\cite{Bask96}. The ``active'' $d_{zx}$ and $d_{yz}$ orbitals
drive the superconducting instability because they have the larger
inter-plane electronic overlap\cite{BJM+00}. This is supported by the
recently implied increase of $T_{\rm c}$ upon uniaxial
pressure\cite{OSM+02} along the crystallographic $c$ axis since the
inter-plane coupling is increased. Such pairing is umklapp scattering
enhanced by the body centered tetragonal lattice\cite{Wern02b}. 

The results of the approach can be summarized as follows. The Cooper
pairs form orbital singlets allowing for an even parity real-space
wavefunction in spite of the spin-triplet configuration. Taking into
account the relatively strong interaction effects in the
system\cite{MRS01} this allows for an almost homogeneous gap function, 
consistent with the experimental
observations\cite{LGL+00,TSN+01,ITY+01,YAM+01}. Any impurity or
defect\cite{MHT+98,MMM99} locally breaks the symmetry of the $d_{zx}$
and $d_{yz}$ orbitals and thus acts as a pair breaker in strict
analogy to magnetic impurities in a spin-singlet
superconductor\cite{AG60}. The quadratic temperature dependence of the
specific heat\cite{NMM00} follows from fluctuations of the internal
degrees of freedom of the order parameter\cite{Wern02b}. On the other
hand, the pair correlations for the $d_{xy}$ electrons are induced by
the interband proximity effect. Since this effect is usually
strong\cite{ZR01} a single gap is assumed leading to consistency with
Andreev reflection experiments\cite{LGL+00}.

In the subspace of the degenerate $d_{zx}$ and $d_{yz}$ orbitals the
possible order parameters can be classified in standard
notation\cite{Volo92} as orbital-singlet spin-triplet components
\begin{equation}\label{Pts}
\langle {P_{t,\mu}^{s}}^{\dagger} \rangle = 
\sum_{\sigma,\sigma',\sigma''\atop \nu,\nu',\bn}
         \sigma^y_{\sigma,\sigma''}\ \sigma^\mu_{\sigma'',\sigma'}\ 
                     \sigma^y_{\nu,\nu'}\ 
  \langle c^{\dagger}_{{\bn},\nu,\sigma}\,
                 c^{\dagger}_{{\bn},\nu',\sigma'}\rangle
\end{equation}
and orbital-triplet spin-singlet components
\begin{equation}\label{Pst}
\langle {P_{s}^{t,\mu}}^{\dagger} \rangle = 
   \sum_{\mu,\nu',\nu''\atop \sigma,\sigma',\bn}
         \sigma^y_{\nu,\nu''}\ \sigma^\mu_{\nu'',\nu'}\ 
                     \sigma^y_{\sigma,\sigma'}\ 
  \langle c^{\dagger}_{{\bn},\nu,\sigma}\,
                 c^{\dagger}_{{\bn},\nu',\sigma'}\rangle .
\end{equation}
Here $\mu=x,y,z$ labels the triplet components, $\sigma^\mu$ are Pauli
matrices, and $c^{\dagger}_{{\bn},\nu,\sigma}$ are the usual electron 
creation operators on site $\bn$ in orbital $d_{zx}$ ($\nu=x$) or
$d_{yz}$ ($\nu=y$) with spin $\sigma$.

   \begin{figure}[bt]
   \epsfxsize=0.5\textwidth
   \epsfclipon
   \centerline{\epsffile{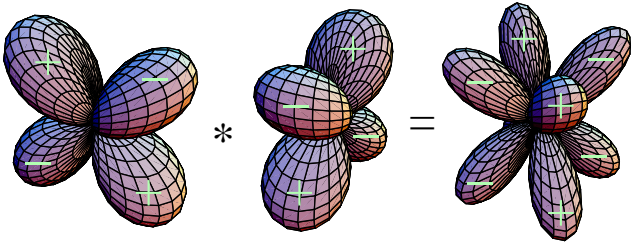}}
   \centerline{\parbox{\textwidth}{\caption{\label{dxyprod}\sl Product
   of the angular components $d_{zx}(\br/r) d_{yz}(\br/r) =
   d_{zx}(\bk/k) d_{yz}(\bk/k)$ in the real or Fourier space
   projection of the pair wave function Eq.\ (\ref{psireal}).}}}      
   \end{figure}

In the presence of Hund's rule coupling the spin-triplet states are
energetically favored over the spin-singlet states and thus form the
ground state of the Cooper pair condensate. Possible effects from 
spin-orbit coupling $\lambda \approx 0.1$ eV\cite{NS00} are
over-compensated by the larger Hund's rule coupling $J_{\rm H} \approx
0.2 - 0.4$ eV\cite{LL00} since $\lambda < J_{\rm H}$\cite{Wern02b}. 

Following Eq.\ (\ref{Pts}) the condensate wavefunction
$|\psi_{\mu}^{s}\rangle = {P_{t,\mu}^{s}}^{\dagger} |0\rangle$ can be
factorized into spatial, spin-triplet, and orbital-singlet
($|x,y\rangle_{s}$) contributions. Since the orbital singlet is odd
and the spin triplet is even under electron exchange the real space
projection of the wave function has to be even. This requirement is
satisfied for electrons in the $d_{zx}$ and $d_{yz}$ orbitals, for
which the total wavefunction is\cite{pwavequote}:
\begin{equation}\label{psireal}
\langle {\bR_\bn + \br} 
|\psi_{\mu}^{s}\rangle = 
                    {d_{zx}(\br)\ d_{yz}(\br)}\
 |x,y\rangle_{s}\
 |\uparrow,\downarrow \rangle_{\mu}\,.
\end{equation}
$\bR_\bn$ are the real-space coordinates of the $\bn^{\rm th}$ Ru ion,
$\br$ are the coordinates relative to $\bR_\bn$. The product of the
angular components $d_{zx}(\br/r) d_{yz}(\br/r)$ has even
parity as shown in Fig.\ \ref{dxyprod}.

The most striking evidence for orbital-singlet pairing in
Sr$_2$RuO$_4$ is the sensitivity of the superconductivity to
impurities\cite{MHT+98} and crystal defects\cite{MMM99}, which can be 
understood by analogy to the effect of magnetic impurities in a
spin-singlet superconductor. In the latter the magnetic impurities 
locally break spin-rotational invariance and thus act as pair
breakers for spin-singlet Cooper pairs\cite{AG60}. Similarly, any
impurity---magnetic, non-magnetic, or crystal defect---locally breaks
the rotational symmetry of the lattice and thus the symmetry between
the $d_{zx}$ and $d_{yz}$ orbitals. Consequently impurities are pair
breaking in the orbital-singlet superconductor described by Eq.\
(\ref{psireal}) in strict analogy to magnetic impurities in a
spin-singlet superconductor. The resulting quantitative applicability
of the theory of Abrikosov and Gor'kov to Sr$_2$RuO$_4$ is
impressively demonstrated in Refs.\ \onlinecite{MHT+98,MMM99}.

As a next step it is necessary to study the superconducting gap
function in order to interpret the numerous directionally dependent
experimental probes. The gap function $\bDelta_\bk = (\Delta_{\bk,x},
\Delta_{\bk,y}, \Delta_{\bk,z})$ is given in the Fourier
representation of Eq.\ (\ref{Pts}) with $\langle
{\bP_{t}^{s}}^{\dagger} \rangle = \sum_\bk \bDelta_\bk^*$ via 
\begin{equation}\label{gapfkt}  
\Delta^*_{\bk,\mu} = 
\sum_{\sigma,\sigma',\sigma''} \sum_{\nu,\nu'}
  \sigma^y_{\sigma,\sigma''}\ \sigma^{\mu}_{\sigma'',\sigma'}\ 
  \sigma^{y}_{\nu,\nu'}\ 
\langle c^{\dagger}_{\bk,\nu,\sigma}\,
        c^{\dagger}_{-\bk,\nu',\sigma'}\rangle .
\end{equation}
It is determined in principle by solving the Eliashberg equations 
\begin{equation}\label{Eliash}  
\sum_{\bk'}  \bDelta_{\bk'}^* = -
\frac{T}{V_{0,2}}\ \boldsymbol{\nabla}_{\!\!\bDelta_\bk}
             \ln \int 
{\cal D}[\phi_i]\ e^{-S_{\rm SG}[\phi_i]}
\end{equation}
self-consistently. $V_{0,2}$ is the effective pairing potential. The
action $S_{\rm SG}[\phi_i]$ has been derived using the quasi
one-dimensionality of the kinetic energy of the $d_{zx}$ and $d_{yz}$
electrons and includes the intermediate coupling on-site interaction
non-perturbatively\cite{Wern02a}. It depends on the four Bose fields,
which can be considered as charge, flavor, spin and spin-flavor fields
in analogy to the two-channel Kondo problem\cite{EK92}, and includes
mass generating terms in the superconducting state\cite{Wern02b}. The
treatment of such a four-component, two-dimensional sine-Gordon action
is quite involved and is only possible using approximations. 

However, to investigate the gap function in Sr$_2$RuO$_4$ it is
sufficient to apply qualitative physical arguments. Starting with the
investigation of the wavefunction symmetry within the non-interacting,
local picture and then analysing the expected influence of strong
interactions it turns out that a rather homogenous gap function must
be expected.  

To establish the wavefunction symmetry in momentum space it is useful
to write\cite{Lind84}
\begin{equation}
\exp\left(i \bk \br \right)= 
 4\pi \sum_{l m} \frac{F_l (kr)}{kr} 
        {Y_{lm}^*(\bk/k)}\ 
i^l\ {Y_{lm}^{\phantom{*}}(\br/r)}\,,
\end{equation}
so that the angular components in real and Fourier space
factorize. $F_l (kr)$ is a regular spherical Bessel function and does
not depend on the magnetization quantum number $m$. Since the $d_{zx}$
and $d_{yz}$ orbitals are linear combinations of the orthogonal
spherical harmonics $Y_{2\,\pm1}$ the angular part of the pair
wavefunction projection onto Fourier space, $\langle {\bk}
|\psi_{\mu}^{s}\rangle$, has the same symmetry as in real space, i.e.,
$d_{zx}(\br/r) d_{yz}(\br/r) = d_{zx}(\bk/k) d_{yz}(\bk/k)$
[Fig.~\ref{dxyprod}]. Introducing the rotation operator ${\cal
  R}_{\pi/2} :\ k_x\to k_y,\ k_y\to -k_x$ one has   
\begin{eqnarray}
{\cal R}_{\pi/2} d_{zx}(\bk/k)
d_{yz}(\bk/k) &=& -d_{zx}(\bk/k) d_{yz}(\bk/k)\,, 
\\
{\cal R}_{\pi/2}
|x,y\rangle_s &=& -|x,y\rangle_s\,,
\end{eqnarray} 
and ${\cal R}_{\pi/2} |\uparrow,\downarrow\rangle_\mu = 
|\uparrow,\downarrow\rangle_\mu$. Consequently 
\begin{equation}
{\cal R}_{\pi/2}\langle
   {\bk} |\psi_{\mu}^{s}\rangle = \langle {\bk}
|\psi_{\mu}^{s}\rangle.
\end{equation}
In other words the wavefunction is even under a rotation of 90$^\circ$
since {\em both} angular and orbital-singlet contributions are odd
under that rotation. We therefore expect the gap function to be of
extended $s$-wave symmetry.

In a group theoretical context the six possible pairing states
described by the pair operators in Eqs.\ (\ref{Pts}) and (\ref{Pst})
find their analogies in the possible pairing states of the tetragonal
point group $D_{4h}$\cite{SAF+99}. Since the angular Fourier space
part $d_{zx}(\bk/k) d_{yz}(\bk/k)$ has even parity and the pair 
wavefunction is invariant under rotation of 90$^{\circ}$ the state
with either $A_{1g}$ or $A_{2g}$ symmetry must be realized. It is 
usefull to define the mirror operators ${\cal M}_x :\ y\to -y$ and
${\cal M}_y :\ x\to -x$ as well as ${\cal M}_\ox :\ \oy\to -\oy$ and
${\cal M}_\oy :\ \ox\to -\ox$ with $\ox=(x+y)/\sqrt{2}$ and
$\oy=(x-y)/\sqrt{2}$. Note that $\ox$ and $\oy$ define a reference
frame rotated by $\pi/4$. Applying these to $d_{zx}(\bk/k)
d_{yz}(\bk/k)$ and $|x,y\rangle_s$ reveals the $A_{2g}$ symmetry of
$\langle {\bk}|\psi_{\mu}^{s}\rangle$.

\begin{figure}[bt]
   \epsfxsize=0.25\textwidth
   \epsfclipon
   \centerline{\epsffile{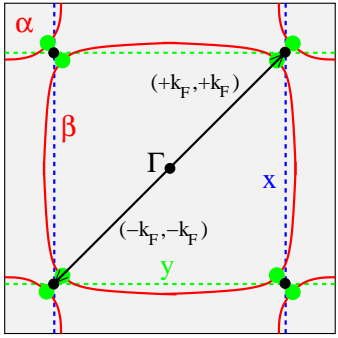}}\vspace{1ex}
   \centerline{\parbox{\textwidth}{\caption{\label{SROfermiBrill}\sl
   Representation of the Fermi surfaces formed by the $d_{zx}$ and
   $d_{yz}$ bands\protect\cite{MS97}. Dashed lines: idealized
   one-dimensional band. Full lines: hybridized bands. Dots: points at
   the Fermi surface where electrons can pair with opposite
   momentum. Small black dots: idealized 1D, larger dots:
   hybridized.}}}      
   \end{figure}

However, in the real system the symmetry of the gap function will be 
significantly altered by hybridization and---more
importantly---interaction effects. Starting out by considering the
non-interacting case the orbital-singlet superconducting instability
can be formulated following BCS\cite{Wern02b} and electrons with
opposite momentum can only pair at the four points
(small black dots in Fig.\ \ref{SROfermiBrill}) of the Brillouin zone
where the idealized, one-dimensional Fermi surfaces\cite{MS97} of the
$d_{zx}$ and $d_{yz}$ bands cross (dashed lines in Fig.\
\ref{SROfermiBrill}). In a more realistic picture the $d_{zx}$ and
$d_{yz}$ bands are weakly hybridized\cite{MS97,Wern02a} and form the
$\alpha$ and $\beta$ sheets of the Fermi surface (full lines in Fig.\
\ref{SROfermiBrill}). It can be shown that then only the eight points
on the $\alpha$ and $\beta$ sheets indicated by the larger dots in
Fig.\ \ref{SROfermiBrill} contribute to the pair formation. Such a
small phase space for the pairing is consistent with the $A_{2g}$
symmetry discussed above but would be inconsistent with the large
specific heat anomaly at the superconducting phase
transition\cite{NMM00}.   

\begin{figure}[bt]
   \epsfxsize=0.44\textwidth
   \epsfclipon
   \centerline{\epsffile{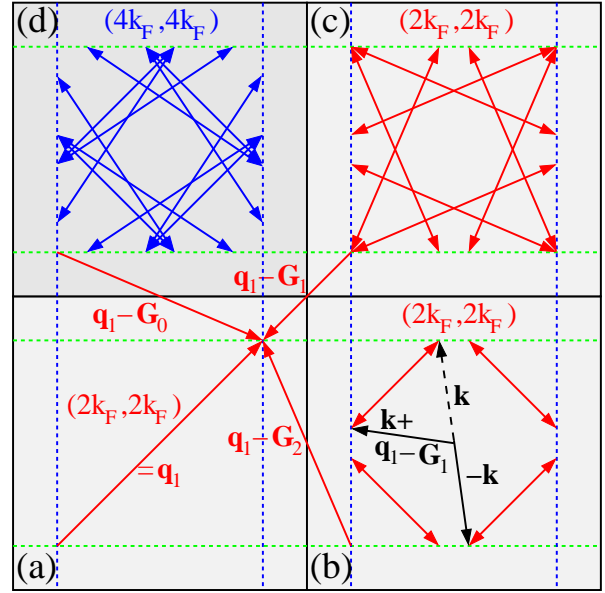}}\vspace{1ex}
   \centerline{\parbox{\textwidth}{\caption{\label{SROpapBrill}\sl
   Representation of the idealized
   one-dimensional Fermi surfaces formed by the $d_{zx}$ and
   $d_{yz}$ bands\protect\cite{MS97} (dashed lines). (a) Magnetic
   momentum transfer $(2k_{\rm F},2k_{\rm F})$ and three
   combinations with reciprocal lattice vectors. (b) and (c) show the
   resulting momentum transfer (double arrows) that allows for mixed
   orbital pairing on many points of the Fermi surfaces. The
   corresponding momenta of the electrons forming a pair are
   illustrated by the black arrows starting from the zone center in
   panel (b). (d) shows the momentum transfer for $(4k_{\rm F},
   4k_{\rm F})$.}}}       
   \end{figure}

That the neglect of the interaction clearly represents an unjustified 
oversimplification of Eq.\ (\ref{Eliash}) becomes obvious from
the significance of the on-site interactions\cite{Wern02a} for the
observed \cite{SBB+99,BSB+02} strong magnetic in-plane
correlations. An estimate of how the interactions increase the pairing
phase space is possible by noting that the dominant magnetic
correlations can be described as gapless, quasi one-dimensional
fluctuations at momentum transfer $\bq_i=(\pm 2k_{\rm F}, \pm 2k_{\rm
  F})$ modulus a reciprocal lattice vector\cite{Wern02a}. The arrows
in Fig.\ \ref{SROpapBrill}(a) show the momentum transfer of $\bq_1 =
(2k_{\rm F}, 2k_{\rm F})$ and three combinations with reciprocal
lattice vectors $\bG_0=(0,2\pi)$, $\bG_1=(2\pi,2\pi)$, and
$\bG_2=(2\pi,0)$ in units of the reciprocal lattice spacing $1/a$.

The back-scattering terms in the action of Eq.\ (\ref{Eliash}) couple
magnetic and charge degrees of freedom\cite{Wern02a}. The Cooper pairs
can thus scatter elastically off the gapless magnetic excitations
modulus any reciprocal lattice vector $\bG_i$, i.e., $\langle
c_{{-\bk},{y},\sigma'} c_{{\bk\pm\bq_i\pm\bG_j},{x},\sigma}
\rangle\neq 0$ as indicated by the black arrows in Fig.\
\ref{SROpapBrill}(b). The resulting momentum transfer allows for mixed
orbital pairing on many points of the Fermi surfaces formed by the
idealized one-dimensional $d_{zx}$ and $d_{yz}$ bands as indicated in
Fig.\ \ref{SROpapBrill}(b) and (c). Including also higher order
contributions allows for an even more homogeneous distribution of
paired electrons across the Fermi surfaces as indicated for $(\pm
4k_{\rm F}, \pm 4k_{\rm F})$ in Fig \ref{SROpapBrill}(d).

This qualitative discussion shows that interactions can be held
accountable for a rather homogeneous gap function in
Sr$_2$RuO$_4$. Moreover, the action $S_{\rm SG}[\phi_i]$ in Eq.\
(\ref{Eliash}) as a function of the charge, flavor, spin, and
spin-flavor Bose fields is manifestly {\em invariant} under the mirror
operations ${\cal M}_\nu$\cite{Wern02a}. The action {\em including}
the interaction thus points towards a $A_{1g}$ symmetry of the gap
function. This results from the fact that the charge, flavor, spin,
and spin-flavor fields are linear combinations of the fields of the
$d_{zx}$ and $d_{yz}$ orbitals. The $A_{1g}$ symmetry is consistent
with thermal conductivity measurements\cite{TSN+01,ITY+01} as well as
with the geometry of the upper critical fields\cite{YAM+01}. Point
contact experiments also do not reveal any significant in-plane
anisotropy of the superconducting order
parameter\cite{LGL+00,Andreevquote}.  

Finally, from the back-scattering terms in the action $S_{\rm 
SG}[\phi_i]$ the existence of two degenerate superconducting saddle
points leading to two degenerate order parameter components can be
deduced\cite{Wern02b}. Each component has a two-fold symmetry
axis. Indeed, the existence of such two order parameter components
with a slight spatial anisotropy in Sr$_2$RuO is implied by the
existence of two upper critical fields\cite{YAM+01,Agte01}. Comparison
with the critical field measurements suggests\cite{Wern02c} that the
components are 93\% isotropic. The two components are classified as
flavor components $\Omega_{{\rm f},x}$ and $\Omega_{{\rm
    f},y}$\cite{Wern02b}.

Since $\Omega_{{\rm f},x}$ and $\Omega_{{\rm f},y}$ are degenerate
in the absence of fields breaking the $\frac{\pi}{2}$-rotational
symmetry the system can fluctuate between the two components in the
ordered phase giving rise to a Goldstone mode\cite{StablequoteL}. This 
mode accounts for\cite{Wern02b} the gapless quasi two-dimensional
excitations observed in the superconducting phase in various
experiments\cite{IMK+00,NMM00,STK+02}. The presence of such a mode in
the superconducting state finds support in the softening of the
in-plane elastic constants recently observed in ultrasonic
measurements\cite{OSM+02}.

In summary the notion of spin-triplet, orbital-singlet pairing in
Sr$_2$RuO$_4$ leads to a straightforward physical picture that is
consistent with fundamental experimental observations such as the
sensivity to impurities, the symmetry of the two upper critical fields
and the fluctuations in the ordered phase. Unlike previous theories
predicting $p$ wave symmetry the two-component order parameter can be
considered as an extended $s$ wave with $A_{1g}$ symmetry and only
slight anisotropy. The interactions play a crucial role in the
in-plane correlations. Details of the non-perturbative approach and
comparisons to experiments as well as the $p$-wave
approach\cite{SAF+99} are given in Refs.\
\onlinecite{Wern02a,Wern02b,Wern02c}.   

I thank M.\ Sigrist as well as D.\ Edwards, M.\ Eschrig, F.\ Laube,
A.\ Rosch, and P.\ Schmitteckert for instructive discussions. I am 
grateful to P.\ C.\ Howell for carefully proofreading the
manuscript. The work was supported by the Center for Functional
Nano\-struc\-tures at the University of Karlsruhe.


\begin{thebibliography}{10}

\bibitem{MYH+97}
Y. Maeno {\it et~al.}, J. Phys. Soc. Jpn. {\bf 66},  1405  (1997).

\bibitem{IMK+00}
K. Ishida {\it et~al.}, Phys. Rev. Lett. {\bf 84},  5387  (2000).

\bibitem{MRS01}
Y. Maeno, T.~M. Rice, and M. Sigrist, Physics Today {\bf 54},  42  (2001).

\bibitem{IMK+98}
K. Ishida {\it et~al.}, Nature (London) {\bf 396},  658  (1998).

\bibitem{LFK+98}
G.~M. Luke {\it et~al.}, Nature (London) {\bf 394},  558  (1998).

\bibitem{DHM+00}
J.~A. Duffy {\it et~al.}, Phys. Rev. Lett. {\bf 85},  5412  (2000).

\bibitem{RS95}
T.~M. Rice and M. Sigrist, J. Phys.: Condens. Matter {\bf 7},  L643  (1995).

\bibitem{LGL+00}
F. Laube {\it et~al.}, Phys. Rev. Lett. {\bf 84},  1595  (2000).

\bibitem{TSN+01}
M.~A. Tanatar {\it et~al.}, Phys. Rev. Lett. {\bf 86},  2649  (2001).

\bibitem{ITY+01}
K. Izawa {\it et~al.}, Phys. Rev. Lett. {\bf 86},  2653  (2001).

\bibitem{SBB+99}
Y. Sidis {\it et~al.}, Phys. Rev. Lett. {\bf 83},  3320  (1999).

\bibitem{MPS00}
I.~I. Mazin, D.~A. Papaconstantopoulos, and D.~J. Singh, Phys. Rev. B {\bf 61},
   5223  (2000).

\bibitem{DLS+00}
A. Damascelli {\it et~al.}, Phys. Rev. Lett. {\bf 85},  5194  (2000).

\bibitem{SDL+01}
K.~M. Shen {\it et~al.}, Phys. Rev. B {\bf 64},  180502(R)  (2001).

\bibitem{NMM00}
S. Nishizaki, Y. Maeno, and Z.~Q. Mao, J. Phys. Soc. Jpn. {\bf 69},  572
  (2000).

\bibitem{STK+02}
M. Suzuki {\it et~al.}, Phys. Rev. Lett. {\bf 88},  227004  (2002).

\bibitem{TK00}
C.~C. Tsuei and J.~R. Kirtley, Rev. Mod. Phys. {\bf 72},  969  (2000).

\bibitem{ZR01}
M.~E. Zhitomirsky and T.~M. Rice, Phys. Rev. Lett. {\bf 87},  057001  (2001).

\bibitem{KS02}
H. Kusunose and M. Sigrist, cont-mat/0205050  (2002).

\bibitem{Bask96}
G. Baskaran, Physica B {\bf 223-224},  490  (1996).

\bibitem{BJM+00}
C. Bergemann {\it et~al.}, Phys. Rev. Lett. {\bf 84},  2662  (2000).

\bibitem{OSM+02}
N. Okuda {\it et~al.}, J. Phys. Soc. Jpn. {\bf 71},  1134  (2002).

\bibitem{Wern02b}
R. Werner, cond-mat/0208307  (2002).

\bibitem{YAM+01}
H. Yaguchi {\it et~al.}, submitted to Phys.\ Rev.\ B  (2002), cond-mat/0106491.

\bibitem{MHT+98}
A.~P. Mackenzie {\it et~al.}, Phys. Rev. Lett. {\bf 80},  161  (1998).

\bibitem{MMM99}
Z. Mao, Y. Mori, and Y. Maeno, Phys. Rev. B {\bf 60},  610  (1999).

\bibitem{AG60}
A.~A. Abrikosov and L.~P. Gor'kov, Zh. Eksp. Teor. Fiz. {\bf 39},  1781
  (1960), [Sov. Phys. JETP {\bf 12}, 1243 (1961)].

\bibitem{Volo92}
G.~E. Volovik, {\em Exotic properties of superfluid $^3$He}, {\em Series in
  Condensed Matter Physics, Vol.1} (World Scientific, Singapore, 1992).

\bibitem{NS00}
K.-K. Ng and M. Sigrist, J. Phys. Soc. Jpn. {\bf 69},  3764  (2000).

\bibitem{LL00}
A. Liebsch and A. Lichtenstein, Phys. Rev. Lett. {\bf 84},  1591  (2000).

\bibitem{pwavequote}
In a $p$-wave superconductor the wavefunction of the electrons in the
  condensate has different parity then the electronic orbital wavefunctions.

\bibitem{Wern02a}
R. Werner and V.~J. Emery, cond-mat/0208306  (2002).

\bibitem{EK92}
V.~J. Emery and S.~A. Kivelson, Phys. Rev. B {\bf 46},  10812  (1992).

\bibitem{Lind84}
A. Lindner, {\em Drehimpulse in der Quantenmechanik} (Teubner, Stuttgart,
  1984).

\bibitem{SAF+99}
M. Sigrist {\it et~al.}, Physica C {\bf 317-318},  134  (1999).

\bibitem{MS97}
I.~I. Mazin and D. Singh, Phys. Rev. Lett. {\bf 79},  733  (1997).

\bibitem{BSB+02}
M. Braden {\it et~al.}, cond-mat/0206304  (2002).

\bibitem{Andreevquote}
The differential resistance data from point contact experiments\cite{LGL+00}
  have been argued to be inconsistent with an isotropic $s$-wave superconductor
  but since a $s$-wave model including gapless excitations might also account
  for the spectra this is not conclusive\cite{Wern02b}.

\bibitem{Agte01}
D.~F. Agterberg, Phys. Rev. B {\bf 64},  052502  (2001).

\bibitem{Wern02c}
R. Werner, cond-mat/0208308  (2002).

\bibitem{StablequoteL}
The tetragonal structure of Sr$_2$RuO$_4$ has been found to be very stable [M.
  Braden {\it et~al.}, Phys. Rev. B {\bf 57}, 1236 (1998)] and consequently the
  degeneracy of the two order parameter components is quite robust.

\end{thebibliography}
\end{document}